\documentclass[journal,onecolumn]{IEEEtran}
\usepackage{amsmath}
\usepackage{booktabs}
\usepackage{latexsym}
\usepackage{mathrsfs}
\usepackage{amsthm}
\usepackage{amssymb}
\usepackage{amsfonts}
\usepackage{amsbsy}

\usepackage[shortlabels]{enumitem}
\usepackage{url}
\usepackage{array}
\usepackage{pdflscape}
\usepackage{xcolor}
\usepackage{stmaryrd}
\usepackage{verbatim}
\usepackage{braket}
\usepackage{graphicx}
\usepackage{tikz}
\newcommand{\Ff}{{\mathbb F}}

\newcommand{\Zz}{{\mathbb Z}}

\newcommand\cc{{\mathcal C}}
\newcommand\qq{{\mathcal Q}}%

\def\tr{\operatorname{tr}}

\theoremstyle{plain}
\newtheorem{thm}{Theorem}
\newtheorem{lem}[thm]{Lemma}
\newtheorem{prop}[thm]{Proposition}

\newtheorem{cor}[thm]{Corollary}
\newtheorem{defn}{Definition}

\def\tr{\operatorname{tr}}
\usepackage[colorlinks=true]{hyperref}

\newcommand\dsb[1]{\llbracket #1 \rrbracket}
\begin{document}
\title{Bounds and Constructions of Quantum Locally Recoverable Codes from Quantum CSS Codes}

\author{Gaojun Luo, Bocong Chen, Martianus Frederic Ezerman and San Ling
\thanks{G. Luo, M. F. Ezerman, and S. Ling are with the School of Physical and Mathematical Sciences, Nanyang Technological University, 21 Nanyang Link, Singapore 637371, e-mails: $\{\rm gaojun.luo, fredezerman, lingsan\}$@ntu.edu.sg.}
\thanks{B. Chen is with the School of Future Technology, South China University of Technology, Guangzhou 510641, China, e-mail: bocongchen@foxmail.com.}
\thanks{G. Luo, M. F. Ezerman, and S. Ling are supported by Nanyang Technological University Research Grant No. 04INS000047C230GRT01. B. Chen is supported in part by the National Natural Science Foundation of China under Grant 11971175, Grant 12171162, and Grant 12371521; and in part by the Guangdong Basic and Applied Basic Research Foundation under Grant 2023A1515010284.}}



\maketitle

\begin{abstract}
Classical locally recoverable codes (LRCs) have become indispensable in distributed storage systems. They provide efficient recovery in terms of localized errors. Quantum LRCs have very recently been introduced for their potential application in quantum data storage. In this paper, we use classical LRCs to investigate quantum LRCs. We prove that the parameters of quantum LRCs are bounded by their classical counterparts. We deduce the bounds on the parameters of quantum LRCs from the bounds on the parameters of the classical ones. We establish a characterization of optimal pure quantum LRCs based on classical codes with specific properties. Using well-crafted classical LRCs as ingredients in the construction of quantum CSS codes, we offer the first construction of several families of optimal pure quantum LRCs.
\end{abstract}


\section{Introduction}\label{sec:intro}

Classical locally recoverable codes (LRCs) were introduced in \cite{Gopalan2012} to repair a single failed node. Such a failure is the most frequent in distributed storage systems. In contrast to maximum distance separable (MDS) codes, classical LRCs provide a very efficient recovery scheme for such a failure, addressing challenges related to fault tolerance, data reliability, and efficient use of resources in distributed storage systems.

Let $q = p^m$ be a prime power, with $p$ being a prime and $m$ a positive integer. Let $\Ff_q$ be the finite field with $q$ elements. We denote by $\Ff_q^n$ the $n$-dimensional space over $\Ff_q$. An $[n,k,d]_q$ linear code $\cc$ is a $k$-dimensional linear subspace of $\Ff_q^n$ with minimum Hamming distance $d$. A classical LRC $\cc$ with \emph{locality} $r$ is a linear code $\cc$ with an additional property related to $r$. For each $i\in\{1,\ldots,n\}$ and each codeword $\mathbf{c}=(c_1,\ldots,c_n) \in \cc$, the $i^{\rm th}$ symbol $c_i$ can be recovered by accessing \emph{at most} $r$ other code symbols of $\mathbf{c}$. The past decade has seen increasingly rapid advances on the topic of classical LRCs. Important bounds on the parameters of such codes have been established in \cite{Prakash12,Cadambe2015,Tamo2016}, and \cite{Hao2020}. Finding explicit constructions of optimal classical LRCs has attracted considerable attention. The results in \cite{Hao2020,Jin2019,Chen2019,Xing2022,Zhang2020,Tamo2015,Chen2018,Qian2020} and \cite{Luo2019} form an important part of our understanding of the topic.

\subsection{Quantum Locally Recoverable Codes}

Very recently, Golowich and Guruswami in \cite{Golowich2023} defined the quantum analogue of LRCs for their potential application in quantum data storage. In quantum locally recoverable process, each codeword is checked by two low-weight stabilizers. In a quantum low-density parity-check (LDPC) code, each codeword must be checked by many low-weight stabilizers. Golowich and Guruswami indicated that quantum LRCs can be identified as a foundational step to study stronger locality properties, such as those strongly suggested by the LDPC property.

A $q$-ary quantum code $\qq$ is a $K$-dimensional subspace of the complex Hilbert space $(\mathbb{C}^q)^{\otimes n}$ of $n$ qudits. The parameters $\dsb{n,\kappa,\delta}_q$ of $\qq$ signifies that it has dimension $q^{\kappa}$ and can correct quantum error operators affecting up to $\lfloor(\delta-1)/2\rfloor$ arbitrary positions in the quantum ensemble. A quantum LRC $\mathcal{Q}$ with locality $r$ is a quantum code with the requirement that, if any single qudit of $\ket{\varphi}\in\mathcal{Q}$ is erased, then each state $\ket{\varphi}\in\mathcal{Q}$ can be recovered by $r$ other qudits of $\ket{\varphi}$ in a recovery channel. The parameters $n,\kappa,\delta,r$ of a quantum LRC satisfy the quantum Singleton-like bound 
\begin{equation}\label{Q-Singleton-dim}
\kappa\leq n-2(\delta-1)-\left\lfloor\frac{n-(\delta-1)}{r+1}\right\rfloor-\left\lfloor\frac{n-2(\delta-1)-\left\lfloor\frac{n-(\delta-1)}{r+1}\right\rfloor}{r+1}\right\rfloor
\end{equation}
given in \cite{Golowich2023}. In the same work, Golowich and Guruswami provided a systematic method to construct quantum LRCs by using (quantum) CSS codes with certain properties. The said authors then presented the first construction of quantum LRCs. The chosen construction route is based on the classical LRCs from the seminal work of Tamo and Barg in \cite{Tamo2014}. 

To the best of our knowledge, it remains open whether the quantum LRCs in \cite{Golowich2023} achieve equality in the bound in \eqref{Q-Singleton-dim}. Hence, there has not been any systematic construction of optimal quantum LRCs in the sense that their parameters meet the bound  with equality.

\subsection{Our Contributions}

At this early stage of research into quantum LRCs, the only known systematic construction is based on the CSS codes, named after Calderbank, Sloane, and Steane. Building on that foundation, this paper focuses on the construction of \emph{optimal} quantum LRCs from CSS codes. The contributions of our investigation can be summarized as follow.
\begin{enumerate}
\item We connect the parameters $n,\kappa,\delta,r$ of a quantum LRC to those of a CSS code with locality $r$. The quantum parameters are constrained by the parameters of the corresponding classical LRCs. Based on the classical bounds, we establish \textbf{two bounds on quantum LRCs}. They are stated formally in \eqref{Q-Singleton} and \eqref{Q-CM}, respectively. They are derived from the Singleton-like bound in \cite[Theorem 2]{Prakash12} and the Cadambe-Mazumdar (CM) bound from \cite{Cadambe2015} on classical LRCs. Asymptotically, the bound in \eqref{Q-CM} is tighter than both the bounds in \eqref{Q-Singleton-dim} and \eqref{Q-Singleton}.

\item We \textbf{characterize} pure quantum LRCs that meet the equality in the bound in \eqref{Q-Singleton}. Let $\cc_1$ and $\cc_2$ be two classical linear codes of length $n$ that satisfy the duality and locality conditions in Proposition \ref{Q-LRC-CON}. We prove that the pure quantum LRC with locality $r$, constructed from $\cc_1$ and $\cc_2$, attains equality in the bound in \eqref{Q-Singleton} if and only if the followings hold:
\begin{itemize}
    \item The codes $\cc_1$ and $\cc_2$ have the same minimum distance and dimension $k$.
    \item Both $\cc_1$ and $\cc_2$ achieve the equality in the bound in \eqref{C-Singleton}, with $\left\lceil\frac{k}{r}\right\rceil= \left\lceil\frac{2k-n}{r}\right\rceil$.
\end{itemize}
\item We discover that \emph{both} bounds in \eqref{Q-Singleton-dim} and \eqref{Q-Singleton} are related to the Singleton bound. Showing that those two bounds are equivalent, however, appears to be challenging. We prove that pure quantum LRCs attain equality in the bound in \eqref{Q-Singleton-dim} if they reach the equality in the bound in \eqref{Q-Singleton}. The advantage of the bound in \eqref{Q-Singleton} is that it provides us with an approach to construct pure quantum LRCs that meet the equality in \emph{both} the bounds in \eqref{Q-Singleton-dim} and \eqref{Q-Singleton}.

\item Based on the characterization in Theorem \ref{pure-optimal}, we present \textbf{several families of optimal pure quantum LRCs} with respect to the bounds in \eqref{Q-Singleton-dim} and \eqref{Q-Singleton}. The constructive technique is to modify known classical LRCs to ensure that they fulfill the conditions required by Proposition \ref{Q-LRC-CON}. These pure quantum LRCs are the first ones that provably meet the equality in the bound in \eqref{Q-Singleton-dim}. Table \ref{table2} list the parameters of the codes.
\end{enumerate}

\begin{table*}[!htbp]
\caption{The parameters of $\dsb{n=u(r+1),\kappa,\delta}_q$ optimal quantum locally recoverable codes with locality $r$}
\label{table2}
\renewcommand{\arraystretch}{1.2}
\centering
\begin{tabular}{cllll}
\toprule
No.  & Distance $\delta$   & Maximal Length $n$ & Constraints & References \\ \midrule
$1$ &    $\delta \in \{2,3,4\}$ & $\Omega(q^2)$ & $r\leq q-1$ and $r>2\delta+u-4$ & Theorems \ref{con-4-2-1} and \ref{con-cyclic-2}  \\

$2$ &    $\delta \in \{5,6\}$  & $\Omega(q^2)$ & $r\leq q-1$ and $r>2\delta+u-4$ & Theorems \ref{con-4-2-1} and \ref{con-cyclic-1}  \\

$3$ &    $\delta \in \{7,8\}$  & $\Omega\left(q^{2-o(1)}\right)$ & $r\leq q-1$ and $r>2\delta+u-4$ & Theorems \ref{con-4-2-1} and \ref{con-cyclic-1}  \\

$4$ &    $\delta \in \{9,10\}$  & $\Omega\left(q^{\frac{3}{2}-o(1)}\right)$ & $r\leq q-1$ and $r>2\delta+u-4$ & Theorems \ref{con-4-2-1} and \ref{con-cyclic-1}  \\

$5$ &    $\delta\geq 11$  & $\Omega\left(q(q\log q)^{\frac{1}{\lfloor(d-3)/2\rfloor}}\right)$ & $r\leq q-1$ and $r>2\delta+u-4$ & Theorems \ref{con-4-2-1} and \ref{con-cyclic-1}  \\
\bottomrule
\end{tabular}
\end{table*}

This paper is organized as follows. Section \ref{sec:pre} reviews useful notions and relevant known results regarding classical and quantum LRCs. In Section \ref{sec:bound}, we formulate and prove bounds on the parameters of quantum LRCs that we construct from CSS codes. We characterize the optimality of the quantum LRCs in terms of the parameters of the ingredient codes in the CSS construction. In Section \ref{sec:codes}, we construct optimal quantum LRCs. Section \ref{sec:con} brings this paper to conclusion and highlights some research directions.

\section{Preliminaries}\label{sec:pre}

Given a positive integer $n$, we use $[n]$ and $[0,n]$ to denote the respective sets $\{1,2,\ldots,n\}$ and $\{0,1,\ldots,n\}$. Let $\cc$ be an $[n,k,d]_q$ linear code. The \textit{dual} code $\cc^{\perp}$ of $\cc$ is
\[
\cc^{\perp}=\left\{(a_1,\ldots,a_{n})\in\Ff_q^n:\sum_{i=0}^{n-1}a_ic_i=0,\
\hbox{for all}\  (c_1,\ldots,c_{n})\in\cc \right\}.
\]
The code $\cc$ is \textit{dual-containing} if $\cc^{\perp}\subseteq \cc$ and is \textit{self-orthogonal} if $\cc\subseteq \cc^{\perp}$. Given a $\mathbf{c}=(c_1,\ldots,c_n)\in\Ff_q^n$, the \textit{Hamming weight} of $\mathbf{c}$ is ${\rm wt}(\mathbf{c})=|\{i:c_i\neq 0\}|$. Let ${\rm wt}(S)$ denote the minimum Hamming weight of the nonzero codewords in $S\subseteq \Ff_q^n$.

\subsection{Classical Locally Recoverable Codes}

We begin by collecting basic results about classical LRCs.

\begin{defn}\label{Definition-LRC}{\rm \cite{Agarwal2018}}
An $[n,k,d]_q$ linear code $\cc$ has locality $r$ if, for each $i\in[n]$ and each codeword $\mathbf{c}=(c_1,\ldots,c_{n})\in\cc$, there exists a linear function $\rho_i:\Ff_q^r\rightarrow \Ff_q$ such that $c_i=\rho_i(c_{t_1},\ldots,c_{t_r})$ with $t_1,\ldots,t_r\in[n]\setminus{i}$.
\end{defn}

Given a codeword $\mathbf{c}=(c_1,\ldots,c_{n})\in\cc$, its {\it support} is ${\rm supp}(\mathbf{c})=\{i\in[n]:c_i\neq 0\}$.

\begin{defn}\label{Definition-LRC-1}
An $[n,k,d]_q$ linear code $\cc$ has locality $r$ if, for each $i\in[n]$, there exists a codeword $\overline{\mathbf{c}}\in\cc^{\perp}$ with ${\rm wt}(\overline{\mathbf{c}})\leq r+1$ and $i\in{\rm supp}(\overline{\mathbf{c}})$.
\end{defn}

Henceforth, unless otherwise stated, all linear codes with locality $r$ coincide with Definition \ref{Definition-LRC-1}. Essential bounds on the parameters of classical LRCs have been established in \cite{Gopalan2012,Prakash12,Tamo2016}, and \cite{Hao2020} by using coding theoretic and combinatorial tools. We reproduce two of them for convenience. It was shown in \cite{Gopalan2012} and \cite{Prakash12} that every $[n,k,d]_q$ linear code with locality $r$ satisfies the Singleton-like bound
\begin{equation}\label{C-Singleton}
d\leq n-k-\left\lceil\frac{k}{r}\right\rceil+2.
\end{equation}
In general, the bound in \eqref{C-Singleton} is not tight, especially for small $q$. Taking the alphabet size of the codes into consideration, Cadambe and Mazumdar in \cite{Cadambe2015} proved that the dimension of each $[n,k,d]_q$ linear code with locality $r$ satisfies what is now known as the CM bound
\begin{equation}\label{C-CM}
k\leq \underset{\ell\in\Zz_{\geq 0}}{\min}\left\{\ell r+k_{opt}^{(q)}(n-\ell(r+1),d)\right\},
\end{equation}
where $\Zz_{\geq 0}$ is the set of all integers $\geq 0$ and $k_{opt}^{(q)}(n,d)$ is the maximum dimension of an $\Ff_q$-linear code of specific length $n$ and minimum distance $d$. Codes with locality $r$ that attain the equality in the bound in \eqref{C-Singleton} or \eqref{C-CM} are said to be \emph{optimal}.

\subsection{Quantum Locally Recoverable Codes}

We use the Dirac notation for quantum mechanics for brevity. Let $\mathbb{C}$ be the field of complex numbers and let $\mathbb{C}^q$ be the $q$-dimensional Hilbert space over $\mathbb{C}$. An \emph{$n$-qudit system} or vector is a nonzero element of $(\mathbb{C}^q)^{\otimes n}\cong\mathbb{C}^{q^n}$. We write $\mathbf{z}=(z_1,\ldots,z_n)\in\Ff_q^n$. The standard basis of $(\mathbb{C}^q)^{\otimes n}$ over $\mathbb{C}$ is
\begin{equation}\label{basis}
\left\{\ket{\mathbf{z}}:=\ket{z_1}\otimes\cdots \otimes\ket{z_n}:\mathbf{z}\in\Ff_q^n\right\}.
\end{equation}
A quantum state of $(\mathbb{C}^q)^{\otimes n}$ is typically written as $\ket{\mathbf{x}}=\sum_{\mathbf{z}\in\Ff_q^n}u_{\mathbf{z}}\ket{\mathbf{z}}$, with $u_{\mathbf{z}}\in\mathbb{C}$ and $\frac{1}{q^n} \sum_{\mathbf{z}\in\Ff_q^n}||u_{\mathbf{z}}||^2=1$. 

The group of \emph{quantum error operators} of $(\mathbb{C}^q)^{\otimes n}$ is
\begin{equation}\label{errorgroup}
\mathcal{E}_n:=\left\{\xi_p^i \, X(\mathbf{a}) \, Z(\mathbf{b}):\mathbf{a},\mathbf{b}\in\Ff_q^n \mbox{ and } i\in\Ff_p\right\}, \mbox{ with } \xi_p=e^{\frac{2\pi\sqrt{-1}}{p}}.
\end{equation}
Letting $\mathbf{a}=(a_1,\ldots,a_n)$ and $\mathbf{b}=(b_1,\ldots,b_n)$, the \textit{support} of an error operator 
\[
E=\xi_p^i \, X(\mathbf{a}) \, Z(\mathbf{b}) \mbox{ is }
{\rm supp}_Q(E)=\{i\in[n]: a_i\neq 0\ \mbox{or}\ b_i\neq 0\}.
\]
The respective actions of $X(\mathbf{a})$ and $Z(\mathbf{b})$ on the basis $\ket{\mathbf{z}}$ in (\ref{basis}) are defined as
\[
X(\mathbf{a})\ket{\mathbf{z}}:=\ket{\mathbf{z}+\mathbf{a}}\ \mbox{and}\ Z(\mathbf{b})\ket{\mathbf{z}}:=\xi_p^{\tr(\langle\mathbf{b},\mathbf{z}\rangle_{{\rm E}})}\ket{\mathbf{z}},
\]
where $\tr(\cdot)$ is the trace function from $\Ff_q$ to $\Ff_p$ and $\langle\mathbf{b},\mathbf{z}\rangle_{{\rm E}}$ is the Euclidean inner product of $\mathbf{b}$ and $\mathbf{z}$.

Formalizing the stabilizer technique, first introduced by Gottesman in \cite{Gottesman1997}, Calderbank, Rains, Shor, and Sloane in \cite{Calderbank1998} proposed a general framework to generate qubit, the shorthand for \emph{quantum bit}, codes by group character theory and finite geometry. Starting with an abelian subgroup $G$ of $\mathcal{E}_n$, the qubit quantum stabilizer code $\mathcal{Q}(G)$ is 
\begin{equation*}
\mathcal{Q}(G)=\left\{\ket{\mathbf{c}}\in(\mathbb{C}^2)^{\otimes n}:E\ket{\mathbf{c}}=\ket{\mathbf{c}},\ \mbox{for any}\ E\in G\right\}.
\end{equation*}
In addition, the group $G$ can be characterized by a classical binary code that is self-orthogonal under the symplectic inner product. This method was subsequently extended to the general qudit, the shorthand for \emph{quantum dit}, {\it i.e.}, the nonbinary $q > 2$ cases, for which a general treatment was supplied by Ketkar, Klappenecker, Kumar, and Sarvepalli in \cite{Ketkar2006}. The CSS codes form a subclass of the stabilizer codes. Some early examples of optimal qubit codes are CSS. They remain popular in numerous small-scale implementation platforms. 

\begin{prop}\label{CSS}{\rm \cite{Ketkar2006}}~{\rm (\textbf{CSS Construction})}
Let $\cc_i$ be an $[n,k_i,d_i]_q$ code for $i \in \{1,2\}$. If $\cc_1^{\perp}\subseteq \cc_2$, then there exists an $\dsb{n,\kappa,\delta}_q$ quantum code with
\[
\kappa=k_1+k_2-n \mbox{ and } 
\delta=\min\left\{ {\rm wt}\left(\cc_2\setminus \cc_1^{\perp}\right), {\rm wt}\left(\cc_1\setminus \cc_2^{\perp}\right)\right\}.
\]
The quantum code is pure whenever $\delta=\min\{d_1,d_2\}$ and is impure if $\delta>\min\{d_1,d_2\}$. 
\end{prop}

From hereon, we set aside the case of $\cc_1^{\perp}=\cc_2$ as the only error operator at work there is the identity matrix. The quantum analogue of Definition \ref{Definition-LRC-1} was formulated in \cite{Golowich2023} by using quantum stabilizer codes.

\begin{defn}\label{Definition-QLRC}{\rm \cite{Golowich2023}}
Let $G$ be an abelian subgroup of the group $\mathcal{E}_n$ of quantum error operators for a fixed $n$. Let 
\[
\mathcal{Q}(G)=\left\{\ket{\mathbf{c}}\in(\mathbb{C}^q)^{\otimes n}:E\ket{\mathbf{c}}=\ket{\mathbf{c}},  \mbox{ for any } E\in G\right\}
\]
be a quantum stabilizer code. The code $\mathcal{Q}(G)$ is locally recoverable with locality $r$ if, for each $i\in[n]$, there exists stabilizers $E_1=X(\mathbf{a}) \, Z(\mathbf{b})$ and $E_2=X(\mathbf{x}) \, Z(\mathbf{y})$ such that $\left|{\rm supp}_Q(E_1)\cup {\rm supp}_Q(E_2)\right|\leq r$,  $(a_i,b_i)=(1,0)$, and $(x_i,y_i)=(0,1)$.
\end{defn}

Definition \ref{Definition-QLRC} enables us to construct quantum stabilizer codes with locality $r$ from classical linear codes.

\begin{prop}\label{Q-LRC-CON}{\rm \cite{Golowich2023}}
Let $\cc_i$ be an $[n,k_i,d_i]_q$ linear code for $i \in \{1,2\}$. Let  
\[
\cc_1^{\perp}\subsetneq \cc_2, \quad \kappa=k_1+k_2-n, \quad 
\delta=\min\{ {\rm wt}\left(\cc_2\setminus \cc_1^{\perp}\right), {\rm wt}\left(\cc_1\setminus \cc_2^{\perp}\right)\}.
\]
If, for each $j\in[n]$, there exist $\overline{\mathbf{c}}_1\in \cc_1^{\perp}$ and $\overline{\mathbf{c}}_2\in \cc_2^{\perp}$ satisfying 
\[
\left|{\rm supp}(\overline{\mathbf{c}}_1)\cup {\rm supp}(\overline{\mathbf{c}}_2)\right| \leq r+1 \mbox{ and } 
j\in {\rm supp}(\overline{\mathbf{c}}_1)\cap {\rm supp}(\overline{\mathbf{c}}_2),
\]
then there exists an $\dsb{n,\kappa,\delta}_q$ quantum code $\mathcal{Q}$ with locality $r$.
\end{prop}

If $\cc_1 = \cc_2$, then the next corollary follows directly from Proposition \ref{Q-LRC-CON}.

\begin{cor}\label{Q-LRC-Cor}
Let $\cc$ be an $[n,k,d]_q$ linear code with $\cc^{\perp}\subsetneq \cc$. Let $\kappa=2k-n$ and $\delta= {\rm wt}\left(\cc\setminus \cc^{\perp}\right)$. If $\cc$ has locality $r$, then there exists an $\dsb{n,\kappa,\delta}_q$ quantum code $\mathcal{Q}$ with locality $r$.
\end{cor}

\section{Bounds on Quantum Locally Recoverable Codes}\label{sec:bound}

In this section, we establish a connection between the parameters of quantum LRCs and those of classical LRCs. We use the connection to derive new bounds on quantum LRCs from the known ones on the corresponding classical codes. Using the Singleton-like bound in \eqref{C-Singleton}, we obtain the quantum Singleton-like bound in \eqref{Q-Singleton}. We characterize the optimality of pure quantum LRCs based on the bound in \eqref{Q-Singleton} by using the corresponding classical LRCs as ingredients in the CSS construction. We conclude that, in terms of their parameters, pure quantum LRCs that achieve the equality in the bound in \eqref{Q-Singleton} also meet the equality in the bound in \eqref{Q-Singleton-dim}.

For brevity, we introduce three quantities on the parameters of a $q$-ary linear code. Let $d_{opt}^q(n,k;r)$ be the \emph{maximum minimum distance} of a code with fixed length $n$, dimension $k$, and locality $r$. Let $k_{opt}^q(n,d;r)$ be the \emph{maximum dimension} of a code with fixed length $n$, minimum distance $d$, and locality $r$. Let $n_{opt}^q(k,d;r)$ be the \emph{minimum length} of a code with fixed dimension $k$, minimum distance $d$, and locality $r$.

\begin{thm}\label{QG-LRC-BOUND}
Let $\mathcal{Q}$ be an $\dsb{n,\kappa=k_1+k_2-n,\delta}_q$ quantum code with locality $r$ constructed from Proposition \ref{Q-LRC-CON}, with $k_1$ and $k_2$ being the respective dimensions of the ingredients $\cc_1$ and $\cc_2$ in the CSS construction. Then we have
\begin{align*}
2 \, \delta & \leq d_{opt}^q(k_1,\kappa;r)+d_{opt}^q(k_2,\kappa;r),\\
2 \, \kappa &\leq k_{opt}^q(k_1,\delta;r)+k_{opt}^q(k_2,\delta;r),\\
n+\kappa & \geq 2\, n_{opt}^q(\kappa,\delta;r).
\end{align*}
\end{thm}

\begin{IEEEproof}
By Proposition \ref{CSS}, there exist two linear codes $\cc_1$ and $\cc_2$ with respective parameters $[n,k_1,d_1]_q$ and $[n,k_2,d_2]_q$ such that $\cc_1^{\perp}\subsetneq \cc_2$.

\begin{enumerate}[wide, itemsep=0pt, leftmargin=0pt, widest={{\bf Case $2$}}]
\item[{\bf Case $1$}:] If ${\rm wt}\left(\cc_2\setminus \cc_1^{\perp}\right) \leq {\rm wt}\left(\cc_1\setminus \cc_2^{\perp}\right)$, then $\delta= {\rm wt}\left(\cc_2\setminus \cc_1^{\perp}\right)$. Let $\mathbf{I}_{g}$ be the identity matrix of order $g$ and let $\mathbf{O}_{g\times h}$ be the $g\times h$ zero matrix. Since $\cc_1^{\perp}\subsetneq \cc_2$, we can express the respective generator matrices of $\cc_1^{\perp}$ and $\cc_2$ as
\[
\begin{pmatrix}
\mathbf{I}_{n-k_1} & \mathbf{G}
\end{pmatrix} \mbox{ and }
\begin{pmatrix}
\mathbf{I}_{n-k_1} & \mathbf{G}\\
\mathbf{O}_{(k_2+k_1-n) \times (n-k_1)} & \mathbf{P}
\end{pmatrix}.
\]
Hence, we know that the matrix $\begin{pmatrix}
\mathbf{O}_{(k_2+k_1-n) \times (n-k_1)} & \mathbf{P}
\end{pmatrix}$ generates an $[n,k_2+k_1-n,d]_q$ code $\cc$. Since all the nonzero codewords in $\cc$ belong to the set $\cc_2\setminus \cc_1^{\perp}$, we have $d \geq {\rm wt}\left(\cc_2 \setminus \cc_1^{\perp}\right)=\delta$. The code $\cc^{\prime}$ generated by $\mathbf{P}$ has parameters $[k_1,k_2+k_1-n, d \geq \delta]_q$. By Proposition \ref{Q-LRC-CON}, the locality of $\cc_2$ is $r$. Since $\cc$ is a strict subset of $\cc_2$, the locality of $\cc$ is at most $r$. This implies that $\cc^{\prime}$ is a $[k_1,k_2+k_1-n,\geq \delta]_q$ LRC with locality at most $r$. Hence, we know that 
\begin{equation}\label{thm-3-1-1}
\delta \leq d \leq d_{opt}^q(k_1, k_1+k_2-n;r), \quad k_1+k_2-n \leq k_{opt}^q(k_1, \delta ;r)\mbox{, and }  k_1 \geq n_{opt}^q(k_1+k_2-n,\delta;r).
\end{equation}
Since the inner product is nondegenerate, $\cc_1^{\perp}\subsetneq \cc_2$ implies $\cc_2^{\perp}\subsetneq \cc_1$. Using the same argument as the one that we had just made, there exists a $\left[k_2,k_1+k_2-n,\geq {\rm wt}\left(\cc_1\setminus \cc_2^{\perp}\right)\right]_q$ LRC with locality $r$. Hence,
\begin{equation}\label{thm-3-1-2}
\delta\leq d\leq d_{opt}^q(k_2, k_1+k_2-n;r), \quad k_1+k_2-n\leq k_{opt}^q(k_2, \delta ;r) \mbox{, and } k_2\geq n_{opt}^q(k_1+k_2-n,\delta;r).
\end{equation}
The desired conclusion follows from \eqref{thm-3-1-1} and \eqref{thm-3-1-2} combined.

\item[{\bf Case $2$}:] If ${\rm wt} \left(\cc_2\setminus \cc_1^{\perp}\right) > {\rm wt} \left(\cc_1\setminus \cc_2^{\perp}\right)$, then the bound in \eqref{Q-Singleton} can be proved by the same method as in the proof of {\bf Case $1$}, with minor adjustments.
\end{enumerate}
\end{IEEEproof}

Theorem \ref{QG-LRC-BOUND} connects the parameters of the quantum to the classical LRCs. It enables us to establish new bounds on quantum LRCs from classical ones in terms of their respective parameters. There are various known bounds on the parameters of classical LRCs, including the Singleton-like bound, the CM bound, and the analogue of the Gilbert-Varshamov bound. They have been established in, {\it e.g.}, \cite{Gopalan2012,Cadambe2015,Tamo2016}, and \cite{Hao2020}. Here we use the Singleton-like and the CM bounds to explicate Theorem \ref{QG-LRC-BOUND} in the next two corollaries.

\begin{cor}\label{Q-LRC-BOUND}
If $\mathcal{Q}$ is an $\dsb{n,\kappa,\delta}_q$ quantum code with locality $r$ constructed from Proposition \ref{Q-LRC-CON}, then
\begin{equation}\label{Q-Singleton}
2 \, \delta \leq n-\kappa-2\left\lceil\frac{\kappa}{r}\right\rceil+4.
\end{equation}
\end{cor}
\begin{IEEEproof}
Theorem \ref{QG-LRC-BOUND} and the bound in \eqref{C-Singleton} lead directly to the bound in \eqref{Q-Singleton}.
\end{IEEEproof}

\begin{cor}\label{CM-LRC-BOUND}
If $\mathcal{Q}$ is an $\dsb{n,\kappa=k_1+k_2-n,\delta}_q$ quantum code with locality $r$ constructed from Proposition \ref{Q-LRC-CON}, with $k_1$ and $k_2$ being the respective dimensions of the ingredient codes in the corresponding CSS construction, then 
\begin{equation}\label{Q-CM}
2\kappa\leq \underset{\ell_1 \in \Zz_{\geq 0}, \, \ell_2\in \Zz_{\geq 0}}{\min}\left \{(\ell_1+\ell_2) \, r + k_{opt}^{(q)} \, (k_1-\ell_1 \, (r+1),\delta)+ k_{opt}^{(q)} \, (k_2-\ell_2(r+1),\delta)\right\}.
\end{equation}
\end{cor}
\begin{IEEEproof}
The bound follows immediately from Theorem \ref{QG-LRC-BOUND} and the bound in \eqref{C-CM}.
\end{IEEEproof}

In general, comparing the bounds in \eqref{Q-Singleton}, \eqref{Q-CM}, and \eqref{Q-Singleton-dim} is not straightforward. The overall picture is easier to grasp in the asymptotic regime, where $n \to \infty$. Given an $\dsb{n,\kappa=k_1+k_2-n,\delta}_q$ quantum code $\mathcal{Q}$, let $R=\frac{\kappa}{n}$ and $\Delta=\frac{\delta}{n}$ be its \textit{rate} and \textit{relative distance}, respectively. Fixing $r$ and making $n \to \infty$ in \eqref{Q-Singleton-dim} and \eqref{Q-Singleton} yield the respective asymptotic formulas 
\begin{align}
\label{I-dim}R&\leq \left(\frac{r}{r+1}\right)^2-\frac{r(2r+1)}{(r+1)^2}\, \Delta + o(1) \mbox{ and}\\
\label{I-distance}R&\leq \frac{r}{r+2}-\frac{2r}{r+2} \, \Delta+ o(1).
\end{align}
A quick inspection confirms that the bound in \eqref{I-dim} is tighter than the bound in \eqref{I-distance}. If $\ell_i=\left\lceil\frac{1}{r+1}\left(k_i-\frac{q}{q-1}(\delta-1)\right)\right\rceil$, for $i\in\{1,2\}$, then $\frac{q-1}{q}\left(k_i-\ell_i(r+1)\right)\leq \delta-1$. Using the Plotkin bound from \cite[Theorem 2.2.1]{huffman2003}, we get
\[
k_{opt}^{(q)} \, (k_1-\ell_1(r+1),\delta)\leq \left\lfloor\log_q\left(\frac{\delta}{r+2}\right)\right\rfloor.
\]
Fixing $r$ and letting $n \to \infty$, the asymptotic form of the bound in \eqref{Q-CM} is
\begin{equation}\label{I-CM}
R\leq \frac{r}{r+2}-\frac{2r}{r+2}\frac{q}{q-1} \, \Delta + o(1).
\end{equation}

\begin{figure}[!ht]
\center
\caption{Comparison of the asymptotic bounds on qubit quantum codes with locality $2$}
\includegraphics[width=0.6\linewidth]{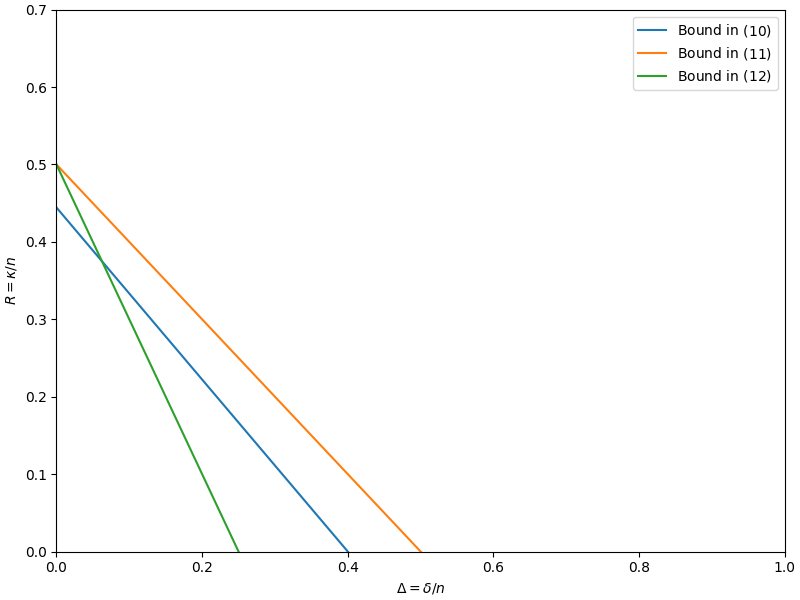}
\label{Fig-1}
\end{figure}

One can quickly verify that the bound in \eqref{I-CM} is tighter than the bound in \eqref{I-distance}. If 
\[
\Delta \geq \frac{q-1}{2q(r+1)^2-(q-1)(r+2)(2r+1)},
\]
then the bound in \eqref{I-CM} is tighter than the bound in \eqref{I-dim}. Given $r=2$ and $q=2$, Figure \ref{Fig-1} presents the comparison of the three bounds in \eqref{I-dim}, \eqref{I-distance}, and \eqref{I-CM} as $n \to \infty$. 

The performance of the bound in \eqref{Q-Singleton-dim} is better than that in \eqref{Q-Singleton} in the asymptotic regime. We are pleased to report that pure quantum codes, with locality $r$, that achieve the equality in the bound in \eqref{Q-Singleton} also attain the equality in the bound in \eqref{Q-Singleton-dim}. To prove this assertion, we characterize a measure of optimality for such codes with respect to the bound in \eqref{Q-Singleton}.

\begin{lem}\label{impure-optimal}
Let $\cc_i$ be an $[n,k_i,d_i]_q$ code, for $i \in \{1,2\}$, that satisfies the conditions in Proposition \ref{Q-LRC-CON}. The pure quantum code with locality $r$, constructed from Proposition \ref{Q-LRC-CON} by using $\cc_1$ and $\cc_2$, cannot achieve the equality in the bound in \eqref{Q-Singleton} whenever $d_1\neq d_2$.
\end{lem}
\begin{IEEEproof}
Without loss of generality, let us assume that $d_1<d_2$. Let $\mathcal{Q}$ be the $\dsb{n,k_1+k_2-n,d_1}_q$ pure quantum code with locality $r$ constructed from Proposition \ref{Q-LRC-CON} by using $\cc_1$ and $\cc_2$. Since $\cc_1$ and $\cc_2$ are codes with locality $r$, by the bound in \eqref{C-Singleton} we get
\begin{equation}\label{thm-2-1}
d_1\leq n-k_1-\left\lceil\frac{k_1}{r}\right\rceil+2 \ \mbox{and}\ d_2\leq n-k_2-\left\lceil\frac{k_2}{r}\right\rceil+2.
\end{equation}
Combining the two inequalities in \eqref{thm-2-1} gives us 
\[
2d_1 <d_1 + d_2 \leq n-(k_1+k_2-n)-\left\lceil\frac{k_1}{r}\right\rceil-\left\lceil\frac{k_2}{r}\right\rceil+4.
\]
Since it is clear that 
\[
\left\lceil\frac{k_1}{r}\right\rceil\geq \left\lceil\frac{k_1+k_2-n}{r}\right\rceil \mbox{ and } 
\left\lceil\frac{k_2}{r}\right\rceil\geq \left\lceil\frac{k_1+k_2-n}{r}\right\rceil,
\]
we arrive at
\begin{equation}\label{thm-2-2}
2d_1<n-(k_1+k_2-n)-\left\lceil\frac{k_1}{r}\right\rceil-\left\lceil\frac{k_2}{r}\right\rceil+4\leq n-(k_1+k_2-n)-2\left\lceil\frac{k_1+k_2-n}{r}\right\rceil+4.
\end{equation}
We can now conclude that $\mathcal{Q}$ can never reach the equality in the bound in \eqref{Q-Singleton}.
\end{IEEEproof}

Based on Lemma \ref{impure-optimal}, we provide a sufficient and necessary condition for pure quantum LRCs to be optimal.

\begin{thm}\label{pure-optimal}
Let $\cc_i$ be an $[n,k_i,d]_q$ code, for $i \in \{1,2\}$, that meets the conditions in Proposition \ref{Q-LRC-CON}. The pure quantum code $\mathcal{Q}$ with locality $r$, constructed from Proposition \ref{Q-LRC-CON} by using $\cc_1$ and $\cc_2$, is optimal with respect to the bound in \eqref{Q-Singleton} if and only if the following two statements hold.
\begin{enumerate}
\item Both $\cc_1$ and $\cc_2$ are optimal with respect to the bound in \eqref{C-Singleton}.
\item $k_1=k_2$ and $\left\lceil\frac{k_1}{r}\right\rceil= \left\lceil\frac{k_1+k_2-n}{r}\right\rceil$.
\end{enumerate}
\end{thm}

\begin{IEEEproof}
Since the sufficient condition follows by the same method as the one used in proving Lemma \ref{impure-optimal}, we only need to provide the proof for the necessary condition.

From the given $\cc_i$s, we know that $\mathcal{Q}$ is an $\dsb{n,k_1+k_2-n,d}_q$ pure code. Since $\mathcal{Q}$ is optimal,
\begin{equation}\label{thm-3-1}
2d= n-k_1+n-k_2-2\left\lceil\frac{k_1+k_2-n}{r}\right\rceil+4.
\end{equation}
Since $\cc_1$ and $\cc_2$ have locality $r$, the bound in \eqref{C-Singleton} implies
\begin{equation}\label{thm-3-2}
d\leq n-k_1-\left\lceil\frac{k_1}{r}\right\rceil+2 \ \mbox{and}\ d\leq n-k_2-\left\lceil\frac{k_2}{r}\right\rceil+2.
\end{equation}
Combining \eqref{thm-3-1} and \eqref{thm-3-2}, we get $2\left\lceil\frac{k_1+k_2-n}{r}\right\rceil \geq \left\lceil\frac{k_1}{r}\right\rceil+\left\lceil\frac{k_2}{r}\right\rceil$ and, since $\left\lceil\frac{k_1}{r}\right\rceil\geq \left\lceil\frac{k_1+k_2-n}{r}\right\rceil$ and $\left\lceil\frac{k_2}{r}\right\rceil\geq \left\lceil\frac{k_1+k_2-n}{r}\right\rceil$, 
\begin{equation}\label{thm-3-3}
2\left\lceil\frac{k_1+k_2-n}{r}\right\rceil = \left\lceil\frac{k_1}{r}\right\rceil+\left\lceil\frac{k_2}{r}\right\rceil.
\end{equation}
Using \eqref{thm-3-1} and \eqref{thm-3-3}, we infer that
\begin{equation*}
2d= n-k_1-\left\lceil\frac{k_1}{r}\right\rceil+2+n-k_2-\left\lceil\frac{k_2}{r}\right\rceil+2.
\end{equation*}
Going back to \eqref{thm-3-2}, we have
\begin{equation}\label{thm-3-4}
d=n-k_1-\left\lceil\frac{k_1}{r}\right\rceil+2=n-k_2-\left\lceil\frac{k_2}{r}\right\rceil+2,
\end{equation}
which is equivalent to 
\[
k_1+\left\lceil\frac{k_1}{r}\right\rceil = k_2+\left\lceil\frac{k_2}{r}\right\rceil \mbox{, forcing } k_1 = k_2.
\]
Using \eqref{thm-3-3}, we conclude that $\left\lceil\frac{k_1}{r}\right\rceil= \left\lceil\frac{k_1+k_2-n}{r}\right\rceil$. The optimality of $\cc_1$ and $\cc_2$ follows from \eqref{thm-3-4}.
\end{IEEEproof}

Now we are ready to prove that the pure code $\mathcal{Q}$ in Theorem \ref{pure-optimal} achieves the equality in the bound in \eqref{Q-Singleton-dim}.

\begin{thm}\label{pure-optimal-1}
Let $\mathcal{Q}$ be a pure quantum code with locality $r$ constructed from Proposition \ref{Q-LRC-CON}. If $\mathcal{Q}$ is optimal with respect to the bound in \eqref{Q-Singleton}, then the code is also optimal with respect to the bound in \eqref{Q-Singleton-dim}.
\end{thm}

\begin{IEEEproof}
By Theorem \ref{pure-optimal}, the optimality of $\mathcal{Q}$ implies the existence of classical codes $\cc_1$ and $\cc_2$ with parameters $[n,k,d]_q$ such that $\left\lceil\frac{k}{r}\right\rceil= \left\lceil\frac{2k-n}{r}\right\rceil$. If $k=vr-x$ and $2k-n=vr-y$ for $0\leq x<r$ and $0\leq y<r$, then $n=vr-2x+y$. Again, by Theorem \ref{pure-optimal}, $\cc_1$ and $\cc_2$ are optimal with regard to the bound in \eqref{C-Singleton}, implying $d=y-x-v+2$. The code $\mathcal{Q}$ has parameters $\dsb{n=vr-2x+y,vr-y,d=y-x-v+2}_q$ and locality $r$. 

Next, we prove that $\mathcal{Q}$ meets the bound in \eqref{Q-Singleton-dim} with equality. Since $0\leq x<r$, 
\begin{align}
n-2(d-1)-\left\lfloor\frac{n-(d-1)}{r+1}\right\rfloor&=vr-y+2v-2-\left\lfloor\frac{vr+v-(x+1)}{r+1}\right\rfloor\nonumber \\
&=vr-y+v-1.\label{thm-3-8-1}
\end{align}
We use the bound in \eqref{Q-Singleton-dim}, the given $0\leq y<r$, and the equality in  \eqref{thm-3-8-1} to obtain 
\begin{align*}
vr-y &\leq n-2(d-1)-\left\lfloor\frac{n-(d-1)}{r+1}\right\rfloor-\left\lfloor\frac{n-2(d-1)-\left\lfloor\frac{n-(d-1)}{r+1}\right\rfloor}{r+1}\right\rfloor\\
&=vr-y+v-1-\left\lfloor\frac{vr-y+v-1}{r+1}\right\rfloor =vr-y
\end{align*}
and, thus, confirm that $\mathcal{Q}$ achieves the bound in \eqref{Q-Singleton-dim} with equality.
\end{IEEEproof}

\section{Constructions of optimal Quantum Locally Recoverable Codes}\label{sec:codes}

This section contains several constructions of optimal quantum LRCs with respect to the bound in \eqref{Q-Singleton}. We follow Theorem \ref{pure-optimal} closely and start with well-known classical LRCs. After proving that they satisfy the conditions in Proposition \ref{Q-LRC-CON}, we build optimal quantum LRCs. We emphasize that, by Theorem \ref{pure-optimal-1}, all quantum LRCs in this section also attain the equality in the bound in \eqref{Q-Singleton-dim}.

\subsection{Constructions from the Parity-Check Matrices of the Classical Codes}

For $n \leq q$, let $\Ff_q^*$ be the multiplicative group of $\Ff_q$ and let $\alpha_1,\ldots,\alpha_n$ be $n$ distinct elements of $\Ff_q$. Let $\mathbf{a}=(\alpha_1,\ldots,\alpha_n)$ and let $\mathbf{v}=(v_1,\ldots,v_n) \in (\Ff_q^*)^n$. Given $k\in[1,n]$, the generalized Reed-Solomon (GRS) code $GRS_k(\mathbf{a},\mathbf{v})$ has a generator matrix 
\begin{equation}\label{eq:GRS}
G_{k,\mathbf{a},\mathbf{v}}=\left(\begin{matrix}
v_1 & v_2 &\cdots & v_n\\
v_1\alpha_1 & v_2\alpha_2 &\cdots & v_n\alpha_n\\
\vdots& \vdots &  \ddots & \vdots\\
v_1\alpha_1^{k-1} & v_2\alpha_2^{k-1} &\cdots & v_n\alpha_n^{k-1}
\end{matrix}\right).
\end{equation}
We know, {\it e.g.}, from \cite[Section 9]{Ling2004}, that $GRS_k(\mathbf{a},\mathbf{v})$ is an $[n,k,n-k+1]_q$ MDS code. The dual code $GRS_{n-k}(\mathbf{a},\mathbf{v}^{\prime})$ of $GRS_k(\mathbf{a},\mathbf{v})$ is a GRS code, for some $\mathbf{v}^{\prime}\in (\Ff_q^*)^n$.

Let $\mathbf{v}_i$ be a vector of length $n_r=r+1$ over $\Ff_q^*$, for each $i\in[u]$. Let $\mathbf{G}_i$ be an $(n_r-k-u)\times (r+1)$ matrix over $\Ff_q$ and let $\mathbf{0}_{r+1}$ be the zero vector of length $r+1$. If a linear code $\cc$ has a parity-check matrix
\begin{equation}\label{LRC-check}
\mathbf{H}=\begin{pmatrix}
\mathbf{v}_1&  \mathbf{0}_{r+1} & \cdots& \mathbf{0}_{r+1}\\
\mathbf{0}_{r+1}&  \mathbf{v}_2 & \cdots& \mathbf{0}_{r+1}\\
\vdots& \vdots &  \ddots & \vdots\\
\mathbf{0}_{r+1}&  \mathbf{0}_{r+1} & \cdots & \mathbf{v}_u\\
\mathbf{G}_1 & \mathbf{G}_2 & \cdots & \mathbf{G}_u
\end{pmatrix},
\end{equation}
then it has locality $r$. In fact, the first $u$ rows of $\mathbf{H}$ ensure that $\cc$ has locality $r$ and the last $n_r-k-u$ rows contribute to the minimum distance of $\cc$. Using parity-check matrices in the form of \eqref{LRC-check}, linear codes with locality $r$ achieving the equality in the bound in \eqref{C-Singleton} were constructed in \cite{Jin2019,Chen2019,Xing2022,Zhang2020}. Their main idea is to use $\mathbf{G}_i\{\mathbf{v}_i\}=\begin{pmatrix} \mathbf{v}_i \\ \mathbf{G}_i\end{pmatrix}$ as a generator matrix $G_{k,\mathbf{a},\mathbf{v}}$ of the GRS code. We reproduce their constructions in the following lemma. Given a vector $\mathbf{a}=(a_1,\ldots,a_{r+1})\in\Ff_q^{r+1}$ with $a_1,\ldots,a_{r+1}$ being distinct elements, we denote by $T_\mathbf{a}=\{a_1,\ldots,a_{r+1}\}$ the collection of all the coordinates of $\mathbf{a}$. 

\begin{lem}[Summary of Constructions in {\rm \cite{Jin2019,Chen2019,Xing2022,Zhang2020}}]\label{Chen}
Let $q$ be a prime power and let $d\geq 2$ be an integer. Let $r$ and $u$ be positive integers such that $r>d-2$ and $r\leq q-1$. For each $i\in[u]$, let $\mathbf{v}_i$ be an arbitrary vector in $\left(\Ff_q^*\right)^{r+1}$ and let $\mathbf{a}_i \in \Ff_q^{r+1}$ be such that its $r+1$ coordinates are distinct. Let $\mathbf{G}_i\{\mathbf{v}_i\}$ be the matrix $G_{d-1,\mathbf{a}_i,\mathbf{v}_i}$ as defined in \eqref{eq:GRS}, for each $i\in[u]$. The code $\cc$ whose parity-check matrix $\mathbf{H}$ is defined as in $\eqref{LRC-check}$ is a $[u(r+1),ur-d+2,d]_q$ optimal LRC with locality $r$ with respect to the bound in \eqref{C-Singleton} if one of the following assertions hold.
\begin{enumerate}
\item For $d \in\{2,3,4\}$, $\mathbf{v}_1=\ldots = \mathbf{v}_u$ and $\mathbf{a}_1=\ldots=\mathbf{a}_u$.
\item For $d\geq 5$, 
\[
\left|\bigcup_{g\in S}T_{\mathbf{a}_g}\right|\geq r \, |S|+1 \mbox{, for each } S \subseteq [u] \mbox{ with } |S|\leq\left\lfloor\frac{d-1}{2}\right\rfloor.
\]
\end{enumerate}
\end{lem}

Following Lemma \ref{Chen}, Table \ref{table1} lists the parameters of some optimal classical LRCs with $d\geq 5$ and locality $r$ from \cite{Jin2019,Chen2019}, and \cite{Xing2022}.

\begin{table*}[!htbp]
\caption{The parameters of an $[n=u(r+1),k,d]_q$ optimal locally recoverable code with locality $r$ and $d-2<r\leq q-1$}
\label{table1}
\renewcommand{\arraystretch}{1.2}
\centering
\begin{tabular}{clll}
\toprule
No.  & Distance $d$   & Length $n$ & References \\ \midrule
$1$ &    $d \in \{2,3,4\}$ & $\infty$ & \cite{Chen2019}  \\

$2$ &    $d \in \{5,6\}$  & $\Omega(q^2)$ & \cite{Jin2019}  \\

$3$ &    $d \in \{7,8\}$  & $\Omega\left(q^{2-o(1)}\right)$ & \cite{Xing2022}  \\

$4$ &    $d \in \{9,10\}$  & $\Omega\left(q^{\frac{3}{2}-o(1)}\right)$ & \cite{Xing2022}  \\

$5$ &    $d\geq 11$  & $\Omega\left(q(q\log q)^{\frac{1}{\lfloor(d-3)/2\rfloor}}\right)$ & \cite{Xing2022}  \\
\bottomrule
\end{tabular}
\end{table*}

We now prove that all optimal LRCs in \cite{Jin2019,Chen2019,Xing2022,Zhang2020} can be used to construct optimal quantum LRCs with respect to the bound in \eqref{Q-Singleton}.

\begin{thm}\label{con-4-2-1}
Let $q$ be a prime power and let $d\geq 2$ be an integer. If $r$ and $u$ are positive integers such that $r>2(d-2)+u$ and $r\leq q-1$, then there exists a $\dsb{u \, (r+1), u \, r-2(d-2)-u,\delta}_q$ pure quantum code $\mathcal{Q}$ with locality $r$. Its length $u \, (r+1)$ and minimum distance $\delta=d$ coincide with the length $n$ and minimum distance $d$ of the corresponding classical code in Table \ref{table1}. The code $\mathcal{Q}$ is optimal with respect to the bound in \eqref{Q-Singleton}.
\end{thm}

\begin{IEEEproof}
We make use of an $[r+1,d-1,r-d+3]_q$ GRS code $GRS_{d-1}(\mathbf{a}_i,\mathbf{v}_i)$ whose dual is $GRS_{r-d+2}(\mathbf{a}_i,\mathbf{v}_i^{\prime})$, for each $i\in[u]$. Since $r>2(d-2)+u$, we know that $GRS_{d-1}(\mathbf{a}_i,\mathbf{v}_i^{\prime})$ is a subcode of $GRS_{r-d+2}(\mathbf{a}_i,\mathbf{v}_i^{\prime})$. Let 
\[
\begin{pmatrix} \mathbf{v}_i \\ \mathbf{G}_i\end{pmatrix} \mbox{ and }
\begin{pmatrix} \mathbf{v}_i^{\prime} \\ \mathbf{G}_i^{\prime}\end{pmatrix}
\]
be the respective generator matrices of $GRS_{d-1}(\mathbf{a}_i,\mathbf{v}_i)$ and $GRS_{d-1}(\mathbf{a}_i,\mathbf{v}_i^{\prime})$, for each $i\in[u]$, making 
\[
\begin{pmatrix} \mathbf{v}_i \\ \mathbf{G}_i\end{pmatrix}
\]
a parity-matrix of $GRS_{r-d+2}(\mathbf{a}_i,\mathbf{v}_i^{\prime})$. Since $GRS_{d-1}(\mathbf{a}_i,\mathbf{v}_i^{\prime})$ is a subcode of $GRS_{r-d+2}(\mathbf{a}_i,\mathbf{v}_i^{\prime})$, we have 
\[
\begin{pmatrix} \mathbf{v}_i \\ \mathbf{G}_i\end{pmatrix}\left(\mathbf{x}\begin{pmatrix} \mathbf{v}_i^{\prime} \\ \mathbf{G}_i^{\prime}\end{pmatrix}\right)^{\top}=\mathbf{0}_{d-1} \mbox{, for each } \mathbf{x}\in\Ff_q^{d-1},
\]
which is equivalent to
\begin{equation}\label{thm-4-2-1-2}
\begin{pmatrix} \mathbf{v}_i \\ \mathbf{G}_i\end{pmatrix}\begin{pmatrix} \mathbf{v}_i^{\prime} \\ \mathbf{G}_i^{\prime}\end{pmatrix}^{\top}=\mathbf{O}_{(d-1)\times(d-1)},\ \mbox{for each}\ i\in[u].
\end{equation}

Using $GRS_{d-1}(\mathbf{a}_i,\mathbf{v}_i)$ and $GRS_{d-1}(\mathbf{a}_i,\mathbf{v}_i^{\prime})$, we generate two codes $\cc_1$ and $\cc_2$ with respective parity-check matrices
\begin{equation}\label{thm-4-2-1-1}
\mathbf{H}=\begin{pmatrix}
\mathbf{v}_1&  \mathbf{0}_{r+1} & \cdots& \mathbf{0}_{r+1}\\
\mathbf{0}_{r+1}&  \mathbf{v}_2 & \cdots& \mathbf{0}_{r+1}\\
\vdots& \vdots &  \ddots & \vdots\\
\mathbf{0}_{r+1}&  \mathbf{0}_{r+1} & \cdots & \mathbf{v}_u\\
\mathbf{G}_1 & \mathbf{G}_2 & \cdots & \mathbf{G}_u
\end{pmatrix}
\ \mbox{and}\
\mathbf{H}^{\prime}=\begin{pmatrix}
\mathbf{v}_1^{\prime}&  \mathbf{0}_{r+1} & \cdots& \mathbf{0}_{r+1}\\
\mathbf{0}_{r+1}&  \mathbf{v}_2^{\prime} & \cdots& \mathbf{0}_{r+1}\\
\vdots& \vdots &  \ddots & \vdots\\
\mathbf{0}_{r+1}&  \mathbf{0}_{r+1} & \cdots & \mathbf{v}_u^{\prime}\\
\mathbf{G}_1^{\prime} & \mathbf{G}_2^{\prime} & \cdots & \mathbf{G}_u^{\prime}
\end{pmatrix}.
\end{equation}
Consulting Table \ref{table1}, $\cc_1$ and $\cc_2$ have parameters $[u(r+1),ur-d+2,d]_q$ and locality $r$, making them optimal with respect to the bound in \eqref{C-Singleton}. By \eqref{thm-4-2-1-1}, for each $j\in[n]$, there exist $\overline{\mathbf{c}}_1\in \cc_1^{\perp}$ and $\overline{\mathbf{c}}_2\in \cc_2^{\perp}$ that satisfy the conditions 
\[
\left|{\rm supp} (\overline{\mathbf{c}}_1)\cup {\rm supp} (\overline{\mathbf{c}}_2)\right|\leq r+1 \mbox{ and } 
j\in {\rm supp}(\overline{\mathbf{c}}_1) \cap {\rm supp}(\overline{\mathbf{c}}_2).
\]
It remains to show that $\cc_2^{\perp}\subseteq \cc_1$. Since, clearly, $\mathbf{H}^{\prime}$ generates the linear code $\cc_2^{\perp}$, seen as a subset of $\cc_2$, we have $\cc_2^{\perp}\subseteq \cc_1$ if and only if 
\[
\mathbf{H} \mathbf{H}^{\prime \top}\mathbf{x}^{\top}=\mathbf{0}_{r-d+2} \mbox{, for each } 
\mathbf{x}\in\Ff_q^{r-d+2}.
\]
We use \eqref{thm-4-2-1-2} to verify that $\mathbf{H}\mathbf{H}^{\prime \top}$ is the zero matrix of order $r-d+2$, confirming that $\cc_2^{\perp}\subseteq \cc_1$.

By Theorem \ref{pure-optimal}, there exists a $\dsb{u(r+1),ur-2(d-2)-u,d}_q$ pure code $\mathcal{Q}$, with locality $r$, that meets the bound in \eqref{Q-Singleton} with equality.
\end{IEEEproof}

\subsection{Constructions from Cyclic Codes}

In this subsection, we use cyclic codes in our constructions of quantum LRCs. Cyclic codes are well-known for having efficient coding and decoding algorithms.

Let $n$ be a positive integer with $\gcd(n,q)=1$. An $[n,k,d]_q$ code $\cc$ is \emph{cyclic} if $(c_{n-1},c_0,\ldots,c_{n-2})\in\cc$ for each codeword $\mathbf{c}=(c_0,c_1,\ldots,c_{n-1})\in\cc$. The codeword $\mathbf{c}$ has a polynomial representation $c(x)=c_0+c_1x+\ldots+c_{n-1} x^{n-1}\in\Ff_q[x]$. We can then identify $\cc$ as an ideal in the residue class ring $R=\Ff_q[x]/\langle x^n-1\rangle$. Since each ideal of $R$ is principal, there exists a unique monic polynomial $g(x)\in R$ of degree $n-k$ such that $\cc=\langle g(x)\rangle$. The polynomial $g(x)$ is the \emph{generator polynomial} of $\cc$ and $h(x)=\frac{x^n-1}{g(x)}$ is its \emph{check polynomial}.

Let $r={\rm ord}_n(q)$ be the smallest positive integer such that $q^r\equiv 1 \pmod{n}$. Let $\alpha$ be a primitive $n^{\rm th}$ root of unity in $\Ff_{q^r}$. The set $D=\{i\in[0,n-1]:g(\alpha^i)=0\}$ is called the \emph{defining set} of the cyclic code $\cc$. The set, seen as the union of some $q$-cyclotomic cosets modulo $n$, is typically defined as $D=\bigcup_{\ell=1}^sC_{i_\ell}$ with $C_{i_\ell}=\{i_\ell q^j:j\in[0,r-1]\}$ being the $q$-cyclotomic coset that contains $i_\ell$. One can quickly check if a vector $\mathbf{w}$ belongs to the cyclic code $\cc$ by using the $s\times n$ matrix
\begin{equation}\label{Cyclic-paritycheck}
H=\begin{pmatrix}
1 & \alpha^{i_1} &\cdots & \alpha^{i_1(n-1)}\\
1 & \alpha^{i_2} &\cdots & \alpha^{i_2(n-1)}\\
\vdots& \vdots &  \ddots & \vdots\\
1 & \alpha^{i_{s}} &\cdots & \alpha^{i_{s}(n-1)}\\
\end{pmatrix},
\end{equation}
since $\mathbf{w} \in \cc$ if and only if $H \, \mathbf{w}^{\top}=\mathbf{0}$. The minimum distance of $\cc$ is $\geq \lambda$ if $D$ contains $\lambda-1$ consecutive integers.

\begin{lem}[BCH bound in \cite{Bose1960,hocquenghem1959codes}]\label{BCH} 
Let $\cc$ be an $[n,k,d]_q$ cyclic code with defining set $D$. Let $m$ be a positive integer such that $\gcd(m,n)=1$ and let $u,b\geq 0$ be integers. If the set $\{u+m(b+i):i\in[0,\lambda-2]\}$ is contained in $D$, then $d\geq \lambda$ .
\end{lem}

Cyclic codes have been instrumental as a construction tool for dual-containing codes. The dual-containing property can be conveniently characterized by the defining set, as shown in \cite[Theorem 4.4.11]{huffman2003}. A cyclic code with defining set $D$ is dual-containing if $D \cap -D = \emptyset$ with $-D=\{-i:i\in D\}$. The works in \cite{Tamo2015,Chen2018,Qian2020,Luo2019} have revealed that cyclic codes also play an important role in the construction of classical LRCs. Luo, Xing, and Yuan proved in \cite{Luo2019} that a cyclic code $\cc$ has locality $r$ if there exists a codeword $\overline{\mathbf{c}}\in\cc^{\perp}$ with ${\rm wt}(\overline{\mathbf{c}})\leq r+1$. It was indicated in \cite{Tamo2015} that a cyclic code with certain defining set has locality $r$. We supply a proof for this claim for completeness.

\begin{lem}[\cite{Tamo2015}]\label{cyclic-LRC}
Let $q$ be a prime power and let $r$ and $u$ be positive integers such that $\gcd(u(r+1),q)=1$. Let $\cc$ be an $[u(r+1),k,d]_q$ cyclic code $\cc$ with defining set $D$. If $D$ contains the subset $\{i(r+1)+1:i\in[0,u-1]\}$, then $\cc$ has locality $r$.
\end{lem}
\begin{IEEEproof}
Let $\mathbf{c}=(c_0,\ldots,c_{n-1})$ be a codeword of $\cc$. Let $r={\rm ord}_{u(r+1)}(q)$ and let $\alpha$ be a primitive $n^{\rm th}$ root of unity in $\Ff_{q^r}$. By \eqref{Cyclic-paritycheck}, we have 
\[
c_0+ \alpha^{i(r+1)+1} \, c_1 + \ldots + \alpha^{(n-1)(i(r+1)+1)} \, c_{n-1}=0 \mbox{, for each } i\in[0,u-1].
\]
Summing up these $u$ equations yields $\sum_{j=0}^{r}\alpha^{ju} \, c_{ju}=0$. Let $\{\theta_1,\ldots,\theta_r\}$ be a basis of $\Ff_{q^r}$ over $\Ff_q$ and let $\alpha^{ju}=\sum_{t=1}^ra_{j,t} \, \theta_t$, for each $j\in[0,r]$. One can then confirmed that 
\[
\sum_{j=0}^{r}a_{i,t}c_{ju}=0 \mbox{, for each } t\in[r].
\]
Since $\alpha^{ju}$ is nonzero for each $j\in[0,r]$, there exists an $\mathbf{a}\in\cc^{\perp}$ with ${\rm supp}(\mathbf{a})\subseteq\{ju:j\in[0,r]\}$. Thus, $\cc$ has locality $r$.
\end{IEEEproof}

We construct a family of quantum LRCs meeting the bound in \eqref{Q-Singleton} with equality from the cyclic codes that reach the equality in the bound in \eqref{C-Singleton}, originally constructed in \cite{Tamo2015}.

\begin{thm}\label{con-cyclic-1}
Let $q$ be a prime power. If $u$, $r$, and $\ell$ are positive integers such that $u+2\ell<r+2$ and $(u r+u)$ divides $(q-1)$, then there exists a $\dsb{u (r+1),u (r-1)-2(\ell-1),\ell+1}_q$ code $\mathcal{Q}$, with locality $r$, that achieves the bound in \eqref{Q-Singleton} with equality.
\end{thm}

\begin{IEEEproof}
Let $A=\{i(r+1)+1:i\in[u-1]\}$ and let $B=[\ell]$. Since $u+2\ell<r+2$, we have $\ell<r$, which further implies that $A$ and $B$ are disjoint. Since $(u r+u)$ divides $(q-1)$, we can let $\cc$ be an $[u (r+1),u r-\ell+1,d]_q$ cyclic code with defining set $A\cup B$. Since the defining set contains $\{i \, (r+1)+1 : i\in[0,u-1]\}$ as a subset, $\cc$ has locality $r$ by Lemma \ref{cyclic-LRC}. Using the bound in \eqref{C-Singleton}, we have $d\leq \ell+1$. By the BCH bound, $d\geq \ell+1$. The code $\cc$ is optimal with respect to the bound in \eqref{C-Singleton}. Since $u+2\ell<r+2$,
\[
\left \lceil\frac{u \, r-\ell+1}{r}\right\rceil= \left\lceil\frac{2(u \, r-\ell+1)-u \, (r+1)}{r}\right\rceil=u.
\]

What remains is to confirm that $\cc^{\perp}\subseteq \cc$. Let 
\begin{equation*}
-A =\{i(r+1)-1:i\in[u-1]\} \mbox{ and } 
-B =\{n-j:j\in[\ell]\}.
\end{equation*}
We quickly verify that $-A\cap A=\emptyset$. Since $2\ell<r-u+2\leq r+1$, we can confirm that $-B\cap B=\emptyset$. Since $\ell<r$, $-A \cap B = -B \cap A=\emptyset$, ensuring that $\cc$ is dual-containing. Corollary \ref{Q-LRC-Cor} and Theorem \ref{pure-optimal} guarantee the existence of a $\dsb{u (r+1),u (r-1)-2(\ell-1),\ell+1}_q$ code $\mathcal{Q}$, with locality $r$, which is optimal with respect to the bound in \eqref{Q-Singleton}.
\end{IEEEproof}

Modifying the constructions in \cite{Luo2019} yields another family of optimal quantum LRCs with respect to the bound in \eqref{Q-Singleton}.

\begin{thm}\label{con-cyclic-2}
Let $q$ be a prime power. If $u$ and $r$ are positive integers such that $u+2<r$, $\gcd(u,q)=1$, $(r+1)$ divides $(q-1)$, and $\gcd(u,q-1)$ divides $\frac{2(q-1)}{r+1}$, then there exists a $\dsb{u \, (r+1),u \, (r-1)-2,3}_q$ code $\mathcal{Q}$, with locality $r$, which is optimal with respect to the bound in \eqref{Q-Singleton}.
\end{thm}
\begin{IEEEproof}
We claim that $A=\{i \, (r+1)+1:i\in[0,u-1]\}$ is the union of some $q$-cyclotomic cosets modulo $u(r+1)$. This is equivalent to saying that, for each $i\in[0,u-1]$, there exists some $j\in[0,u-1]$ such that
\begin{equation}\label{thm-4-con-2-1}
q(1+i \, (r+1)) \equiv 1+ j \, (r+1) \pmod{u(r+1)}.
\end{equation}
Since $(r+1)$ divides $(q-1)$, the congruence $(r+1) \,x \equiv q-1\pmod{u \, (r+1)}$ has exactly $r+1$ incongruent solutions, which implies the existence of a $j\in[0,u-1]$ as in \eqref{thm-4-con-2-1}. This settles our claim of $A$ being the union of some $q$-cyclotomic cosets. Since $\gcd(u,q-1)$ divides $\frac{2(q-1)}{r+1}$, the congruence 
\[
q(x(r+1)+2) \equiv x(r+1)+2 \pmod{u(r+1)}
\]
has a solution $y\in[0,u-1]$, ensuring that $B=\{y \, (r+1)+2\}$ is a $q$-cyclotomic coset modulo $u \, (r+1)$.

Using the same argument as in the proof of Theorem \ref{con-cyclic-1}, the cyclic code whose defining set is $A\cup B$ leads to the code $\mathcal{Q}$ with locality $r$ and parameters $\dsb{u (r+1),u (r-1)-2,3}_q$. The code $\mathcal{Q}$ is optimal with respect to the bound in \eqref{Q-Singleton}.
\end{IEEEproof}

\section{Concluding remarks}\label{sec:con}
We have investigated quantum LRCs and provided several constructions based on Proposition \ref{Q-LRC-CON}, which is the only known systematic approach for the construction of quantum LRCs to date. Using the bounds on the parameters of classical LRCs, we have established the corresponding bounds for quantum LRCs. Asymptotically, one of our bounds is tighter than the Singleton-like bound in \eqref{Q-Singleton-dim}. Based on the Singleton-like bound for the classical LRCs, we have proved the analogous bound for quantum LRCs, which is stated in \eqref{Q-Singleton}. We characterize the optimality of pure quantum LRCs with respect to the bound in \eqref{Q-Singleton}. We have shown in Theorem \ref{pure-optimal-1} that pure quantum LRCs that meet the equality in the bound in \eqref{Q-Singleton} also attain the bound in \eqref{Q-Singleton-dim} with equality. By modifying known classical LRCs, we have constructed several families of pure quantum LRCs meeting both bounds in \eqref{Q-Singleton-dim} and \eqref{Q-Singleton} with equality.

Quantum LRCs form a new research topic in quantum error control. They combine local recoverability and quantum coding. Our investigation leads us to some interesting questions.
\begin{itemize}
\item Can we use methods, other than the CSS approach, to construct quantum LRCs via classical codes? Examples of such methods include a construction based on the stabilizer formalism, known as the Hermitian construction. The main ingredients are Hermitian self-orthogonal or dual-containing classical codes. The CSS construction can, in fact, be seen as a special case of the Hermitian construction. The latter is essential as it produces quantum MDS codes with flexible parameters and longer lengths than the CSS construction can cover. Proving that the Hermitian construction can lead to  new quantum LRCs will certainly be beneficial.
\item It is well-known that all quantum MDS codes are pure. Are there \emph{impure} quantum LRC that can meet the bound in either \eqref{Q-Singleton-dim} or \eqref{Q-Singleton} with equality?
\item Do quantum codes with entanglement admit local recoverable property?
\end{itemize}

\end{document}